\documentclass[epj,final]{svjour}
\usepackage{graphics}
\begin{document}
\hyphenation{Coul-omb ei-gen-val-ue ei-gen-func-tion Ha-mil-to-ni-an
  trans-ver-sal mo-men-tum re-nor-ma-li-zed mas-ses sym-me-tri-za-tion
  dis-cre-ti-za-tion dia-go-na-li-za-tion in-ter-val pro-ba-bi-li-ty
  ha-dro-nic he-li-ci-ty Yu-ka-wa con-si-de-ra-tions spec-tra
  spec-trum cor-res-pond-ing-ly}
\title{On technically solving an effective QCD-Hamiltonian}
\headnote{    Preprint MPIH-V11-1999    }
\author{
        Susanne Bielefeld
      , Jan Ihmels
        \thanks{now: Pembroke College, Cambridge CB21RF}
, and Hans-Christian Pauli
       }
 \offprints{\emph{Prof. H.C. Pauli, MPI f\"ur Kernphysik,
         Postfach 10 39 80, D-69029 Heidelberg. },\\
  archive:  hep-th/9900xxx
  }
\institute{Max-Planck-Institut f\"ur Kernphysik, Heidelberg.
           \email{S.Bielefeld@mpi-hd.mpg.de}
           }
\date{06 april 1999}
\abstract{
  By their very nature, field-theoretical Hamiltonians are
  derived in momentum representation.
  To solve the corresponding integro-differential equations 
  is more difficult than to solve the simpler differential
  equations in configuration space (`Schr\"odinger equation').
  For the latter many different and very effective methods 
  have been developed in the past.
  But rather than to Fourier-transform to configuration space~--
  which is not always easy --~the equations are solved here
  directly in momentum space, by using Gaussian quadratures.
  Special attention is given to the case where the
  potential in configuration space is linear and where
  the corresponding momentum-space kernel has an almost
  intractable $1/(\vec k - \vec k ^\prime)^4$-singularity.
  Its regularization requires a certain technical effort, 
  introducing suitable counter terms. 
  The method is numerically reliable and fast, 
  faster than other methods in the literature.
  It should be useful to and also applicable in other approaches, 
  including phenomenological Schr\"odinger-type equations.
} 
\authorrunning{S.Bielefeld et.al}
\titlerunning{On technically solving an effective Hamiltonian for QCD}
\maketitle
%
\section{Introduction}
\label{introduction}
Dirac's front form of Hamiltonian dynamics \cite{dirac49} 
seems to be a good candidate for addressing to the thus far unsolved 
bound-state problem of a (gauge) field theory, 
particularly of non-Abelian quantum chromo-dynamics (QCD),
as reviewed recently in \cite{bpp97}.
Advantages are its comparatively simple vacuum structure and
the simple boost properties.
In fact, one can formulate the bound-state problem frame-independently, 
and if one uses the light-cone gauge $A^{+}=0$, the vacuum is trivial.
More recently \cite{pau98}, the front-form Hamiltonian for QCD
has been reduced to a renormalized, effective Hamiltonian which acts 
only in the Fock space of one quark ($q$) and one anti-quark ($\bar{q}$). 
One then faces the numerical problem to solve these equations.

In principle, one could proceed like in \cite{kpw92,trittmann} for QED. 
But we prefer to move on slowly, by first suppressing in 
Section~\ref{motivation}
all fine and hyperfine interactions with a simple trick.
The resulting spin-less equation finds its analogue
in the `central potential' of a Schr\"odinger equation.
In Section~\ref{lin_pot}, the so truncated effective interaction 
is shown to interpolate smoothly~--
as a function of only one numerical parameter~-- 
between a pure Coulomb and a linear potential.
These potentials in configuration space are presented in detail, 
together with the known analytical solutions to the Coulomb 
and the linear potential.
Their knowledge is advantageous to check the numerical procedures.

In Section~\ref{gaussian} the counter term technology
is presented in detail, and applied in Sections~\ref{coulomb} and
\ref{yukawa} to the Coulomb and the Yukawa, and in Section~\ref{combi} 
to the combined problem, respectively.
The respective eigenvalues and eigenfunctions are presented
in all detail. More and perhaps redundant numerical results
are compiled in the Appendix.
A summary and a discussion in Section~\ref{discussion} 
rounds-off the paper.
\section{The light-cone momentum-space approach}
\label{motivation}
The front-form Hamiltonian for QCD has been reduced in \cite{pau98}
to an effective Hamiltonian, with a resulting
integral equation in momentum space
\begin{eqnarray} 
\lefteqn{
     M^2\langle x,\vec k_{\!\perp}; \lambda_{q},
    \lambda_{\bar q}  \vert \psi\rangle} &&\nonumber\\ & = & 
    \left[ 
    {\overline m^{\,2}_{q} + \vec k_{\!\perp}^{\,2}\over x } +
    {\overline m^{\,2}_{\bar q} + \vec k_{\!\perp}^{\,2}\over 1-x}   
    \right]\langle x,\vec k_{\!\perp}; \lambda_{q},
    \lambda_{\bar q}  \vert \psi\rangle \nonumber\\ 
    & & -{1\over 3\pi^2}
    \sum _{ \lambda_q^\prime,\lambda_{\bar q}^\prime}
    \!\int \frac{ dx^\prime d^2 \vec k_{\!\perp}^\prime }
    { \sqrt{ x(1-x) x^\prime(1-x^\prime)} } 
    {\overline\alpha(Q) \over Q  ^2}
    \nonumber\\ & & 
    \langle
    \lambda_q,\lambda_{\bar q}\vert S(Q)\vert 
    \lambda_q^\prime,\lambda_{\bar q}^\prime\rangle
    \,\langle x^\prime,\vec k_{\!\perp}^\prime; 
    \lambda_q^\prime,\lambda_{\bar q}^\prime  
    \vert \psi\rangle
~.\label{eq:i65}
\end{eqnarray}
The spectrum of the invariant-mass squared eigenvalues
$M ^2$ is the goal we want tot reach. The corresponding eigenfunctions
$\vert \psi\rangle$ are 
the probability amplitudes for finding a quark 
with longitudinal momentum fraction $x$,
transversal momentum $\vec k_{\!\perp}$, 
and helicity $\lambda_{q}$, and correspondingly for the anti-quark.
The physical (renormalized) masses of the quarks are denoted by 
$\overline m _{q}$.
The 4-momentum transfer along the quark line 
$Q _{q}^2=-(k-k')^{\mu}(k-k')_{\mu}$ is generally different from $Q
_{\bar q} ^2$. 
Their mean $Q ^2= (Q _{q} ^2+Q _{\bar q} ^2)/2$
appears in Eq.(\ref{eq:i65}).
The (renormalized) `running' coupling constant 
$\overline\alpha(Q)$ is a function 
of the momentum transfer and will be given below.
The  spinor factor 
\begin{eqnarray}
 \lefteqn{\langle
  \lambda_{q},\lambda_{\bar q}\vert S(Q)\vert 
  \lambda_{q}^\prime,\lambda_{\bar q}^\prime\rangle}&&\\ &=&
 {\left[ \overline u (k_q,\lambda_q)\gamma^\mu
   u(k_q^\prime,\lambda_q^\prime)\right]}\, 
     {\left[ \overline v(k_{\bar q},\lambda_{\bar q}) 
    \gamma_\mu 
   v(k_{\bar q}^\prime,\lambda_{\bar q}^\prime)\right]}
\nonumber\end{eqnarray} 
represents the current-current coupling and describes all fine and
hyperfine interactions. 

To reduce the complexity of the problem we introduce several
simplifications and approximations, as follows. First, we 
suppress all fine and hyperfine interactions by setting
\begin{eqnarray}
\label{current}
{\left[\overline u (k_q,\lambda_q)\gamma^\mu
   u(k_q^\prime,\lambda_q^\prime)\right]}\, 
     {\left[ \overline v (k_{\bar q},\lambda_{\bar q}) 
    \gamma_\mu 
   v(k_{\bar q}^\prime,\lambda_{\bar q}^\prime)\right]}
\left|_{x=\frac{1}{2};\vec{k}_{\perp}=0}\right. && \\ 
\simeq 4 \overline m _q \overline m _{\bar{q}}\,\delta_{\lambda_q
   \lambda_q^\prime}\,\delta_{\lambda_{\bar q}\lambda_{\bar
   q}^\prime}&&\nonumber \ .
\end{eqnarray} 
The helicity summations in Eq.(\ref{eq:i65}) therefore drop out,
and with equal quark masses $\overline{m}_q =\overline{m}_{\bar{q}}=m$ 
the equation simplifies to
\begin{eqnarray}
   M^2 \psi(x,\vec{k}_{\!\perp})&=&\frac{m^2 +
   \vec{k}_{\!\perp}^2}{x(1-x)}\psi(x,\vec{k}_{\!\perp})\\
   &-&\frac{m^2}{\pi^2}\!\int\!
   \frac{ dx^\prime d^2 \vec k_{\!\perp}^\prime\, 
          \psi(x',\vec k_{\!\perp}^\prime)}
        { \sqrt{ x(1-x) x^\prime(1-x^\prime)} } 
   \,\frac{4}{3}\frac{\bar{\alpha}(Q)}{Q^2}
\nonumber\ .
\label{eq:mass} \end{eqnarray}
The longitudinal momentum fraction ($0 \leq x \leq 1$)
and the transversal momentum
($-\infty \leq k_{\perp,x} \leq \infty$) have different domains of
integration. It is convenient \cite{kpw92,trittmann} 
to transform integration variables 
$(x,\vec{k}_{\perp})\,\to\,(k_z,\vec{k}_{\perp})$ by 
\begin{equation}
    x=x(k_z)=\frac{1}{2}\left(1+\frac{k_z}{\sqrt{m^2
    +\vec{k}\,^2_{\perp}+k^2_z}}\right)
\ .\end{equation}
All components of the `vector' 
$\vec{k}=(k_z,\vec{k}_{\perp})$ have then the same domain. 
The corresponding Jacobian is
\begin{equation}
   \frac {dx}{x(1-x)}= \frac {2}{m} 
   \,\frac {dk_z} {A ^2(x,\vec{k}_{\!\perp})}
\ ,\end{equation}
with the $A$-factor defined by
\begin{equation}
   m ^2 A ^4(x,\vec{k}_{\!\perp}) = 
   \frac {m^2+ \vec {k} _{\!\perp}^{\,2} } {4x(1-x)} =
   m^2+ \vec {k} _{\!\perp}^{\,2} + k_{z}^{2} 
\ .\end{equation}
The wave function 
$\psi$ can always be substituted by an other function $\phi$, 
\begin{equation}
   \psi(x,\vec{k}_{\!\perp})=
   \frac{A(x,\vec{k}_{\!\perp})}{\sqrt{x(1-x)}} 
   \phi (x,\vec{k}_{\!\perp})
\ .\end{equation}
Replacing the invariant mass-square
eigenvalue $M^2$  by the binding energy $E$, {\it i.e.} 
\begin{equation}
   M^2=4m^2\,+4mE
\ ,\end{equation}
the original integral equation Eq.(\ref{eq:i65}) becomes finally
\begin{equation}
   \left[E-\frac{\vec{k}^2}{m}\right]\phi(\vec{k})=-\frac{1}{2\pi^2}\int
   \frac {d^3\vec{k'}} {A(\vec k)A(\vec k ^\prime)}\,
   \frac{4}{3}\frac{\bar{\alpha}(Q)}{Q^2}\,\phi(\vec{k'})
\label{start}\ .
\end{equation}
Note that the only approximation is the suppression 
of the fine and hyperfine interaction by Eq.(\ref{current}).

At this point it is completely irrelevant that the
equation holds actually for the front form.
One simply does not recognize its origin.
It could be an equation in the instant form,
where only the (usual) three momenta $\vec k $ play
the role of integration variables.
It would however not be trivial to Fourier transform this
equation to configuration space, the factor $A$
is preventing us to do that in a standard way.
We therefore apply one further approximation.
Somewhat doubtfully, we apply the non-relativistic 
approximation under the integral,
and set 
\begin{equation}
   A(\vec k)\simeq A(\vec k ^\prime)= 1,
   \quad
   Q^2 =(\vec k - \vec k ^\prime)^2
\ .\end{equation}
One now has a complete analogy with the simpler spin-less case 
of an average potential. One could argue
in favor of this approximation,  that it 
is only consistent with Eq.(\ref{current}). 
By replacing the QCD running coupling $4/3\ \bar{\alpha}(Q)$ 
with the (physical) QED coupling $\alpha$ one obtains the usual 
Coulomb-Schr\"odinger equation in momentum space. 

Amazingly enough, just by redefining the wave functions, 
we have derived an equation whose kernel is 
manifestly rotationally invariant. This is surprising since
rotations perpendicular to the $z$-axis are dynamic operators 
in the front form and very complicated \cite{bpp97}.

What shall be used for the effective coupling constant $\bar{\alpha}(Q)$? 
In \cite{pau98} an explicit expression was given, but
here we use the simplified expression of Brodsky {\it et al.}~\cite{bj97}
\begin{equation}
\bar{\alpha}(Q)=\frac{12\pi}{25\ln{\left[\frac{\mu^2
    +Q^2}{\kappa^2}\right]}} =
    \frac{12\pi}{25\ln{\frac{\mu^2}{\kappa^2}}+25\ln{\left[1 +
    \frac{Q^2}{\mu^2}\right]}}
\label{coupl_approx}\ ,\end{equation}
and adopt their numerical values
\begin{equation}
   \kappa \simeq  160\,\,\mbox{MeV}~,\hspace{2em}\mu \simeq  872\,\,\mbox{MeV}
\label{numbers}\ .\end{equation}
Now the problem is set. 

Approximating the logarithm 
$\ln{(1 + Q^2/ \mu^2)}$ by $Q^2/ \mu^2$ we parameterize the kernel of
  integral equation~(\ref{start}) as  
\begin{equation} 
   \frac{4}{3}\frac{\bar{\alpha}(Q)}{Q^2}=
   \frac{2 s^2}
   {(\vec k - \vec k ^\prime) ^2(c^2+ (\vec k - \vec k ^\prime)^2)}
\label{eff.const}\ .\end{equation}
With the parameter values of Eq.(\ref{numbers}) one has 
\begin{equation}
\label{s.c.def}
s = 1070\,\,\mbox{MeV}~,\hspace{2em}c = 1605\,\,\mbox{MeV}~.
\end{equation}
Note that for $c\to 0$, the coupling constant in Eq.(\ref{eff.const})
exposes a $1/(\vec k - \vec k ^\prime) ^4$ singularity. 

\section{The potential in configuration space}
\label{lin_pot}
\begin{figure}
\resizebox{0.48\textwidth}{!}{
 \includegraphics{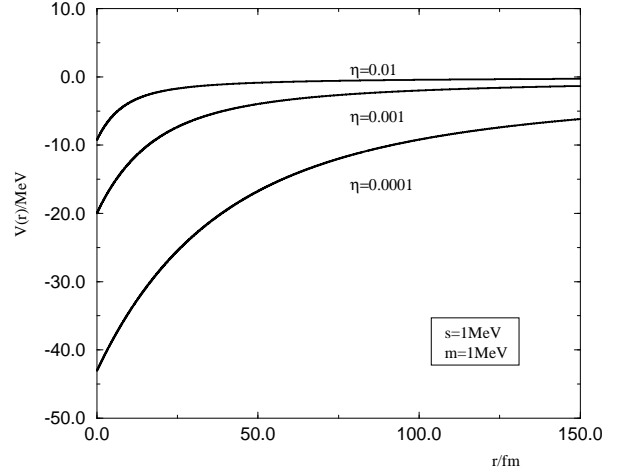}}
\caption{\label{fig:pot}
   The potential $V(r)$  is plotted versus $r$ for different values of
   $\eta=2c/\beta m$.-- Note that for $r \to 0$ the potential goes to
   the finite value $-2s^2/c$, with a linear slope.}
\end{figure}

\begin{figure}
\resizebox{0.48\textwidth}{!}{
 \includegraphics{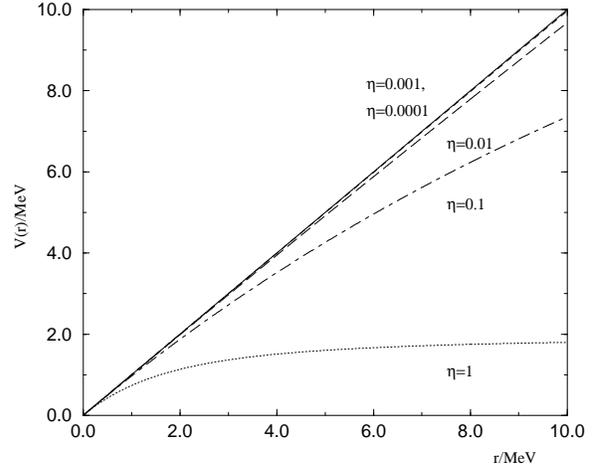}}
\caption{\label{fig:1}
   Same as Fig.~\protect{\ref{fig:pot}}, but with the potential
  normalized to $V(0)=0$.} 
\end{figure}

Before solving numerically the equation
\begin{eqnarray}
\label{eqn_to_solve}
\left[E-\frac{\vec{k}^2}{m}\right]\phi(\vec{k})=-\frac{s^2}{\pi^2}\int
    d^3\vec{k'}\,\frac{1}{Q^2(Q^2+c^2)}\,\phi(\vec{k'})&&\\
\mbox{with}\hspace{2em}Q^2=(\vec{k}-\vec{k'})^2 &&~,\nonumber 
\end{eqnarray}
we discuss its structure by transformation to
configuration space. 
Applying the Fourier transformations according to
\begin{eqnarray}
\label{eq:regpot1}
\psi(\vec{x})&=&\int d^3\vec{k}\,e^{i\vec{k}\vec{x}}\,\phi(\vec{k})~,
\nonumber \\
V({\vec{x}})&=&\int d^3\vec{q}\,e^{i\vec{q}\vec{x}}\,U(\vec{k},\vec{k'})~,\\
U(\vec{k},\vec{k'})&\equiv &
-\frac{s^2}{\pi^2}\frac{1}{(\vec{k}-\vec{k'})^2((\vec{k}-\vec{k'})^2 +c^2)}~,\nonumber 
\end{eqnarray}
one obtains a Schr\"odinger equation
\begin{equation}
\label{ft_E}
\left[E+\frac{\nabla^2}{2m_r}\right]\psi(\vec{x})=V(\vec{x})\,\psi(\vec{x})~~,
\end{equation}
with the reduced mass $m_r=m/2$.
The simple structure is a consequence of the kernel
$U(\vec{k},\vec{k'})$ depending only on the difference $\vec{k}-\vec{k'}$.
Since $\int
d^3\vec{q}\,e^{i\vec{q}\vec{x}}\,1/(Q^2+c^2)=-2\pi^2\,e^{-cr}/r$, with
$r=\vert\vec{x}\vert$, one has
\begin{equation}
V({\vec{x}})=\frac{\beta}{r}\left[e^{-cr}- 1\right]
\end{equation}
where $\beta=2 s^2/c^2$ is similar to the usual fine-structure
constant.
The potential is a superposition of a Coulomb and a Yukawa potential, as
visualized in Figs.~\ref{fig:pot} and~\ref{fig:1}. The behavior 
\begin{equation}
\label{eq:ftregpot}
 V(\vert\vec{x}\vert)=\left\{
 \begin{array}{ll}
   -\frac{2 s^2}{c} + s^2 r & \mbox{for}\,\, r \to 0~, \\
   -\frac{2 s^2}{c^2}\,\frac{1}{r} & \mbox{for}\,\, r \to \infty~,
 \end{array}
\right.
\end{equation}
can be observed in the figures:
For sufficiently small distances the potential is linear
(up to an additional constant), and for large $r$ it
becomes a Coulomb potential. 
For either of these extremes the analytic solutions are known.

The eigenvalue $E$ used in Eq.(\ref{eqn_to_solve}) depends on the
 three parameters $m$, $s$, and $c$. It is easy to show that it
 depends on them in the dimensionless combination 
\begin{equation}
\label{eta_def}
\eta=\frac{c^3}{m s^2}=\frac{2}{\beta}\,\frac{c}{m}~.
\end{equation}
By introducing the dimensionless variables
\begin{equation}
  p=\frac{k}{m_r \beta}~,\hspace{1em} q^2=
  \frac{Q^2}{m_r^2\beta^2}~,\hspace{1em} \epsilon=\frac{2 E}{m_r\beta^2} 
\end{equation}
we obtain
\begin{equation}
\label{eps_eta}
\left[\epsilon-p^2\right]\phi(p)=-\frac{1}{\pi^2}
\int d^3 p'\,\left[\frac{1}{q^2}-\frac{1}{q^2+\eta^2}\right]\,\phi(p')
\end{equation}
which is much simpler than Eq.(\ref{eqn_to_solve}), and we will
therefore focus on this version.

It has the two solvable limiting cases:
(1.) For  $\eta \to \infty$, the equation essentially shows the $q^{-2}$
singularity of a Coulomb problem. (2.) For $\eta \to 0$, the equation
has a $q^{-4}$ singularity corresponding to a linear potential, whose
eigenfunctions are the Airy functions.     

The eigenvalues for the Coulomb problem are thus in our units
\begin{equation}
\epsilon_{n} = -\,\frac{1}{n^2}~,\hspace{1em}n=1,2,\ldots ~.
\end{equation}
The eigenvalues for the linear potential are
\begin{equation}
\label{bareps_behave}
\epsilon_{n}=-2 \eta + \xi_n \eta^{\frac{4}{3}}~,\hspace{1em}n=1,2,\ldots ~.
\end{equation}
where $\xi_n$ are the zeros of the Airy functions $Ai(\xi_n)=0$.

The knowledge of these two limiting cases is very useful for testing
the numerical results.

\section{Numerical solutions by Gaussian quadratures}
\label{gaussian}
Eq.(\ref{eps_eta}) is an integral equation in the three variables
$\vec{p}$. Restricting to s-waves one can integrate out the angles, 
which leads to an equation in one variable $p=\vert \vec{p}
\vert$,~{\it i.e.}
\begin{eqnarray}
\label{eqn_solve_Bohr}
\left[\epsilon-p^2\right]\phi(p)&=&\frac{1}{\pi}
\int_{0}^{\infty}
dp'\,\frac{p'}{p}\,\left[\ln\frac{(p-p')^2}{(p+p')^2}\right. \\
&&-\,\left.\ln
 \frac{(p-p')^2+\eta^2}{(p+p')^{2}+\eta^2}\right]\,\phi(p')~.\nonumber
\end{eqnarray}
It is convenient~\cite{ellerp89}, to convert this integral equation for
the unknown function $\phi(p)$ to a matrix equation for the unknown
numbers $u_i\equiv\sqrt{\omega_i}\phi_i$ with $\phi_i \equiv \phi(p_i)$, 
\begin{equation}
\label{eq:eval}
 (p_i\,^{2}+\,a_{ii})\,u_i + \sum_{j\neq i} a_{ij}u_j=\epsilon\,
 u_i ~,\quad ~i,j=1,\ldots N,
\end{equation}
namely by approximating the integral with the
finite sum of Gaussian quadratures,  
\begin{equation}
  \label{eq:Gauss-Leg}
  \int_{0}^{\infty} f(p) dp \longrightarrow \sum_{i=1}^{N} \omega_i f(p_i)~~.
\end{equation}
The non-diagonal matrix elements are given by 
\begin{eqnarray}
  \label{eq:matnondiag}
 a_{ij}=\frac{1}{\pi}\sqrt{\omega_i\omega_j}\frac{p_j}{p_i}
\left[\ln{\frac{(p_i-p_j)^2}{(p_i+p_j)^2}}\right.&&\\
&&-\left.\ln{\frac{(p_i-p_j)^2+\eta^2}{(p_i+p_j)^2+\eta^2}}\right]~.\nonumber
\end{eqnarray}
It is numerically convenient to map the infinite interval of
Eq.(\ref{eqn_solve_Bohr}) onto a finite one by transforming
variables~\cite{wdipl}, according to
\begin{equation}
  \label{trans}
  \int_0^{\infty}f(p)dp=\int_{-1}^1 f(p(y))\frac{dp}{dy}dy~~~.
\end{equation}
The mapping function $y(p)$ is arbitrary, but must satisfy the
boundary conditions
\begin{equation}
  \label{eq:yofp}
  y(p=0)=-1~, \hspace{2em} y(p=\infty)=+1 ~.
\end{equation}
We choose
\begin{equation}
  \label{eq:trans1}
  y(p)=2e^{-\frac{p}{z}}-1~~~,
\end{equation}
with an adjustable 'stretching' parameter $z$. With $c(y)\equiv
dp/dy$, the quadratures then become  
\begin{equation}
  \label{eq:newquad}
  \sum_i f(p_i)\omega_i=\sum_i f(y_i) c(y_i) \hat{\omega}_i ~~~,
\end{equation}
{\it i.e.} the transformation changes the weights $\omega_i$ into
$\omega_i\equiv c(y_i)\hat{\omega}_i$.  Thus
\begin{equation}
  \label{eq:trans2}
  p_i=-z\,\ln\frac{1+y_i}{2} \hspace{2em} \mbox{ and } \hspace{2em}
  \omega_i=\hat{\omega}_i\frac{z}{1+y_i}~~,
\end{equation}
where the $\hat{\omega}_i$ and $y_i$ are the tabulated weights and
abscissas for the interval $[-1,1]$.  
Finally Eq.(\ref{eq:matnondiag}) can be solved by conventional matrix
diagonalization methods. 

Eq.(\ref{eq:eval}) poses a problem: The diagonal matrix
elements $a_{ii}$ diverge logarithmically. Problems like these can be
solved by the Nystr{\o}m method - by the technique of counter
terms - as to be discussed next~\cite{wdipl}. In principle one adds
and subtracts in Eq.(\ref{eps_eta}) a diagonal term $F(p)\phi(p)$. The
idea is that one of them is treated analytically and the other by Gaussian
quadratures, such that the singularity in $a_{ii}$ cancels. 

The construction of suitable counter terms must be done separately for
the Coulomb and the Yukawa part. Therefore we discuss these two cases
in Sections~\ref{coulomb} and~\ref{yukawa} individually. In
Section~\ref{combi} we return to the full problem.

\section{The Coulomb problem}
\label{coulomb}
The Coulomb problem in momentum space was treated by
W\"olz~\cite{wdipl,kpw92} as a numerical exercise and will be repeated
briefly. In analogy to Eq.(\ref{eqn_solve_Bohr}) we want to solve
\begin{equation}
  \label{eq:coul}
  \left[\epsilon -
  p^2\right]\phi(p)=\frac{1}{\pi}\int_0^{\infty}dp'\frac{p'}{p} 
  \ln\left[\frac{(p-p')^2}{(p+p')^2}\right]\phi(p')~. 
\end{equation}
Adding and subtracting the analytically integrable Coulomb counter
term $F_{C}(p)\phi(p)$ 
\begin{figure}
\resizebox{0.48\textwidth}{!}{
\includegraphics{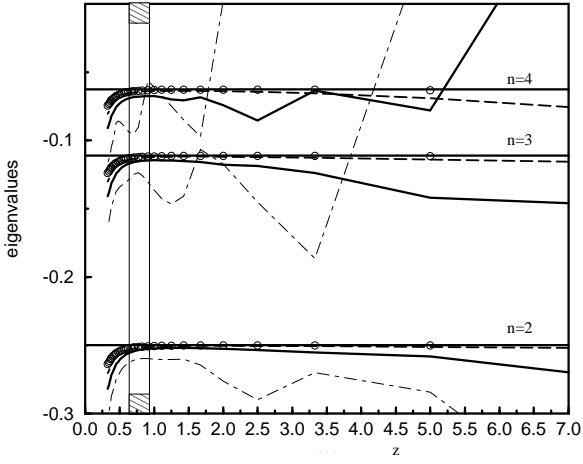}}
\caption{\label{fig:test_coul} 
   The Bohr eigenvalues for $n=2,3,4$ are plotted versus the
   stretching factor $z$ for different matrix dimensions $N$
   ($-\cdot-$ N=8, \protect\rule{5mm}{0.5mm} N=16,
   \protect\rule{3mm}{0.5mm} \protect\rule{3mm}{0.5mm} N=32 and
   $\circ$ N=64).- The exact eigenvalues $-1/n^2$ are shown as
   well.- Note the hatched area, in which the numerical results are
   particularly stable against the number $z$.}
\end{figure} 
\begin{eqnarray}
  \label{eq:counter}
  F_{C}(\vec{p})&=&\frac{1}{\pi^2}\int
  d^3\vec{p}\,'\frac{1}{(\vec{p}-\vec{p}\,')^2}\frac{(1+\vec{p}\,^2)^2}{(1+\vec{p}\,'^
2)^2}\\
 &=&1+\vec{p}\,^2 ~.
\end{eqnarray}
one arrives at
\begin{eqnarray}
  \label{eq:coulG}
 \lefteqn{\left(\epsilon-p^2 + (1+p^2)\right)\,\phi(p)=}\\\nonumber
& & \frac{1}{\pi}\int_0^{\infty}dp'
 \frac{p'}{p}\,\ln
 \biggl[\frac{(p'-p)^2}{(p'+p)^2}\biggr]\left[\phi(p')- 
 \frac{(1+p^2)^2}{(1+p'^2)^2}\phi(p)\right]~.
\end{eqnarray}
For $p=p'$, the term in the square bracket vanishes, {\it i.e.} it is
justified to restrict to $j \neq i$ in the summation over $j$ on the
r.h.s. of Eq.(\ref{eq:eval}).
The diagonal matrix elements then become
\begin{equation}
\label{matrixfin1}
a_{ii}=-(1+p_i^2)-\frac{1}{\pi}\sum_{j\neq i}\omega_j
  \frac{p_j}{p_i}\ln\frac{(p_i-p_j)^2}{(p_i+p_j)^2}\frac{(1+p_i^2)^2}{(1+p_j^2)^2}~.     
\end{equation}
The off-diagonal matrix-elements in Eq.(\ref{eq:matnondiag}) are not
affected by this procedure  
\begin{equation}
  \label{eq:finmat1}
   a_{ij}=\frac{1}{\pi}\sqrt{\omega_i
   \omega_j}\frac{p_j}{p_i}\,\ln\frac{(p_i-p_j)^2}{(p_i+p_j)^2} ~,
\end{equation}
see Eq.(\ref{eq:matnondiag}).
In the sequel we
ask ourselves whether this results can be improved by a stretching
factor $z$, as given in Eq.(\ref{eq:trans2}). The advantage of the
stretching factor is that a particular choice shifts the bulk of
integration points to the region where the wave function is significantly
different from zero. In Fig.~\ref{fig:test_coul} the numerical eigenvalues of
Eq.(\ref{eq:eval}) with the matrix elements of
Eqs.(\ref{matrixfin1}) and~(\ref{eq:finmat1}) are plotted versus $z$ for
different matrix dimensions $N$. As seen in the
figure, the functions $\epsilon_n(z)$ are rapidly varying (almost
fluctuating) for low dimensionality and become flatter with increasing
$N$. For a $z$ within the hatched area, however, the numerical results
are rather stable as function of $N$. A value of $N=32$ (or $16$) seems to
satisfy all practical requirements. The lowest eigenvalue
($\epsilon_1=-1.0$) is not shown, since the function ($\epsilon_1(z)$) 
is completely flat. These observations are somewhat more quantified in
Table~\ref{couleval}. 
\begin{table*}[t]
\begin{center}
\begin{tabular}{|l||c||c|c|c||c|c|c|}\hline
&Exact&\multicolumn{6}{|c|}{Calculated}\\ \hline\hline
n&$-1/n^2$&    N=16   & N=32 & N=64 & N=16   & N=32 & N=64 \\\hline
& &\multicolumn{3}{|c||}{$z=0.70$}&\multicolumn{3}{|c|}{$z=1.0$}\\
\hline\hline 
1&-1.0000&-1.0000  &-1.0000 &-1.0000 &-1.0000 &-1.0000 &-1.0000\\
2&-0.2500&-0.2545  &-0.2525 &-0.2516 &-0.2523 &-0.2510 &-0.2506\\ 
3&-0.1111&-0.1160  &-0.1135 &-0.1126 &-0.1145 &-0.1122 &-0.1117\\ 
4&-0.0625&-0.0683  &-0.0649 &-0.0639 &-0.0676 &-0.0637 &-0.0630\\
5&-0.0400&-0.0471  &-0.0425 &-0.0414 &-0.0485 &-0.0415 &-0.0406\\\hline
\end{tabular}
\caption{\label{couleval}The eigenvalues of the Coulomb problem for
  two values of the stretching factor $z$ and three
  matrix dimensions $N$ are compared with the exact values.}
\end{center}
\end{table*}

\begin{figure}[ht]
\resizebox{0.48\textwidth}{!}{
 \includegraphics{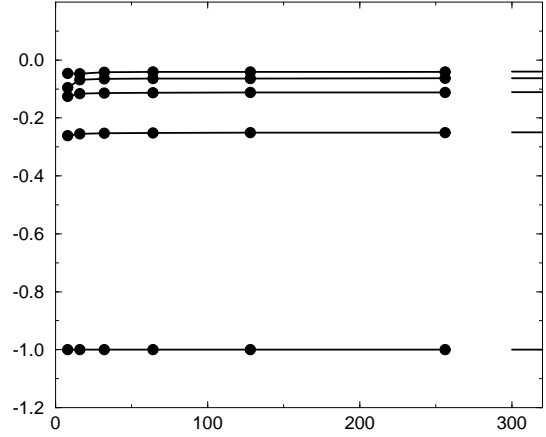}}
\caption{\label{fig:2} 
   The numerical eigenvalues for the Coulomb problem are plotted versus the  
   number of integration points $N$ ($8,16,32,64,128,256$) for
   $z=0.70$.} 
\end{figure}

\begin{figure}[!ht]
\resizebox{0.48\textwidth}{!}{
\includegraphics{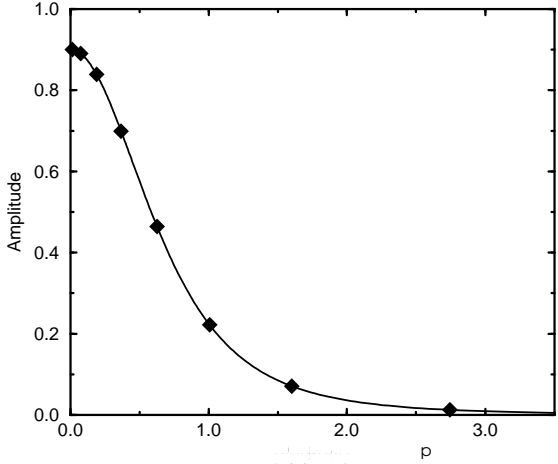}}
\caption{\label{fig:3}
   The exact wave function $\phi(p_i)_{1s}$ for the Coulomb problem
   is plotted versus $p$ (the momentum in units of the Bohr momentum),
   and compared with the numerical values 
   $u_i\sqrt{4\pi\omega_i}$ (filled diamonds). Parameter values are
   $z=0.70$ and $N=8$.- 
   Note the excellent agreement.}  
\end{figure}

The final results for the low lying part of the spectrum are
accumulated in Fig.~\ref{fig:2}. The spectrum is remarkably
insensitive to the matrix dimensions $N$. Also the numerical
wave function as displayed in Fig.~\ref{fig:3} is highly accurate.
For all practical needs $N=16$ is sufficient, see also Table~\ref{couleval}.

One concludes that the particular choice of the stretching factor
($z=0.70$) can be important for improving the rate of convergence and
the precision of the Gaussian method for solving equations in momentum
space.

\section{The Yukawa potential}
\label{yukawa}
\begin{figure}[ht!]
\resizebox{0.48\textwidth}{!}{   
\includegraphics{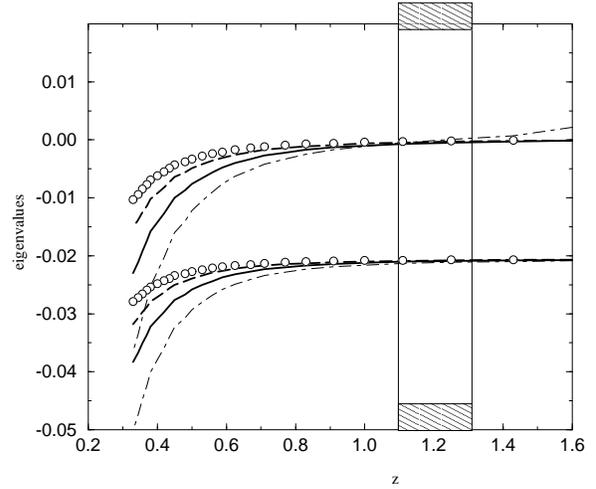}}
\caption{\label{fig:test_yuk} 
   The two lowest eigenvalues of the Yukawa problem with $\eta=1.0$
   are plotted versus the stretching factor $z$ for different matrix
   dimensions: ($-\cdot-$ N=8, \protect\rule{5mm}{0.5mm} N=16,
   \protect\rule{3mm}{0.5mm} \protect\rule{3mm}{0.5mm} N=32 and
   $\circ$ N=64).}
\end{figure}

The Yukawa potential in momentum space obeys the integral equation
\begin{equation}
  \label{eq:intB}
  \left[\epsilon -
  p^2\right]\phi(p)=\frac{1}{\pi}\int_0^{\infty}dp'\frac{p'}{p} 
  \ln\left[\frac{(p-p')^2+\eta^2}{(p+p')^2+\eta^2}\right]\phi(p')~. 
\end{equation}
In analogy to the Coulomb problem, one adds and subtracts an
analytically integrable counter term
$F_{Y}(p)\phi(p)$ 
\begin{eqnarray}
  \label{eq:counteryuk}
  F_{Y}(\vec{p})&=&\frac{1}{\pi^2}\int
  d^3\vec{p}'\frac{1}{(\vec{p}-\vec{p}')^2
  +\eta^2}\frac{(1+\vec{p}\,^2)^2}{(1+\vec{p}\,'^2)^2}\\
  &=&\frac{(1+\vec{p}\,^2)^2(\vec{p}\,^2+(\eta-1)^2)}{(\vec{p}\,^2 +
  \eta^2 -1)^2 +4\vec{p}\,^2} ~.\nonumber 
\end{eqnarray}
The integral equation (\ref{eq:intB}) is then rewritten as
\begin{eqnarray}
  \label{eq:Komplett}
 \lefteqn{\left(\epsilon-p^2 + \frac{(1+p^2)^2 (p^2 +
     (\eta-1)^2)}{(p^2+\eta^2-1)^2 +4p^2}  \right)\,\phi(p)=}\nonumber\\
& & \frac{1}{\pi}\int_0^{\infty}dp'
 \frac{p'}{p}\,\ln
 \biggl[\frac{(p'-p)^2+\eta^2}{(p'+p)^2+\eta^2}\biggr]\\ 
&\times&\left[ \phi(p')-\frac{(1+p^2)^2}{(1+p'^2)^2}\phi(p)\right]~.\nonumber 
\end{eqnarray}
The matrix equation (\ref{eq:eval}) has then the diagonal elements 
\begin{eqnarray}
a_{ii}&=&-\frac{(p_i^2+(\eta-1)^2)(1+p_i^2)^2}{(p_i^2+\eta^2-1)^2+4p_i^2}\\
&&-\frac{1}{\pi}\sum_{j\neq
  i}\omega_j\frac{p_j}{p_i}\ln\biggl[\frac{(p_j-p_i)^2+\eta^2}{(p_j+p_i)^2+\eta^2}\biggr]\frac{(1+p_i^2)^2}{(1+p_j^2)^2}~,\nonumber  
\end{eqnarray}
while the off-diagonal elements are not modified
\begin{equation}
  \label{eq:finmat2}
   a_{ij}=\frac{1}{\pi}\sqrt{\omega_i
   \omega_j}\ln\frac{(p_j-p_i)^2+\eta^2}{(p_j+p_i)^2+\eta^2}~. 
\end{equation}
The integral over the domain $[0,\,\scriptstyle{\infty}]$ is mapped
on the interval $[-1,1]$ as given in Eqs.(\ref{trans}) to (\ref{eq:trans2}).
The limit $\eta \to 0$ gives back the Coulomb problem, see
Section~\ref{coulomb}. 
We have checked explicitly that the programs reproduce this case. A
comparatively small value of $\eta$ is $\eta=0.01$. Indeed, the low
lying part of the spectrum and the wave function 
is very similar to the corresponding result for the
Coulomb problem in Figs.~\ref{fig:2} and~\ref{fig:3}. In either case,
the eigenvalues are practically insensitive to $N$, particular for the
stretching parameter $z=0.7$, chosen to be the same as for the Coulomb
problem. One should note that the Yukawa problem has only a finite
number of bound states. The value of $z=1.25$ is
obtained by the same optimalization procedure and shown explicitly in
Fig.~\ref{fig:test_yuk}. Finally we summarize our best values for
$\eta=0.01$ and $\eta=1.0$ in Table~\ref{yukeval_1_001}. 
\begin{table}[ht!]
\begin{center}
\begin{tabular}{|l||c|c||}\hline
n&  $\eta=0.01$   & $\eta=1.0$ \\
 &  $z=0.7$       & $z=1.25$   \\\hline\hline 
1& -0.8141 & -0.0208 \\
2& -0.1022 & -0.0003 \\
3& -0.0084 &  --     \\
4& -0.0019 &  --     \\
5& -0.0019 &  --     \\\hline
\end{tabular}
\caption{\label{yukeval_1_001}The eigenvalues of all bounded s-states
 for the Yukawa problem for two values of $\eta$. The matrix dimension
 is either $N=32$.}
\end{center}
\end{table}

\section{The hadronic Coulomb potential}
\label{combi}
Combining appropriately the considerations for the Coulomb and the
Yukawa problem by adding and subtracting the counter terms
$(F_{C}(p)+F_{Y}(p))\phi(p)$ as given in Sections~\ref{coulomb}
and~\ref{yukawa}, yield immediately the improved diagonal elements 
\begin{eqnarray}
a_{ii}&=&-(1+p_i^2)+2\eta+\frac{(p_i^2+(\eta-1)^2)(1+p_i^2)^2}{(p_i^2+\eta^2-1)^2+4p_i^2}\\     
&&-\frac{(1+p_i^2)^2}{\pi p_i}\sum_{j\neq i}\omega_j
     p_j\ln\left[\frac{(p_i-p_j)^2}{(p_i+p_j)^2}\right]\frac{1}{(1+p_j^2)^2}\nonumber\\  
&&+\frac{(1+p_i^2)^2}{\pi p_i}\sum_{j\neq
     i}\omega_j p_j
   \ln\left[\frac{(p_j-p_i)^2+\eta^2}{(p_j+p_i)^2+\eta^2}\right]\frac{1}{(1+p_j^2)^2}\nonumber ~,  
\end{eqnarray}
which together with the unchanged off-diagonal elements $a_{ij}$ as
given in Eq.(\ref{eq:matnondiag}) define the matrix equation
(\ref{eq:eval}). 

\begin{figure}
\resizebox{0.48\textwidth}{!}{
 \includegraphics{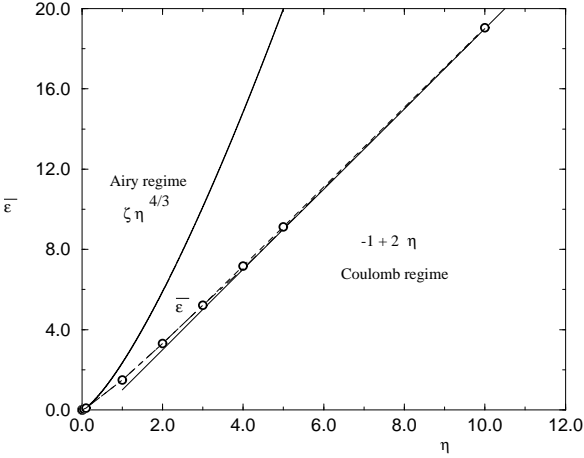}}
\caption{\label{fig:test_g1}The lowest eigenvalue
  $\bar{\epsilon}(\eta)=\epsilon + 2 \eta$ is
  plotted versus $\eta$ ($\circ$). The upper solid
  line indicates the Airy-type solution and the lower solid line
  visualizes the Coulomb-type solution.}
\end{figure}

In the limit $\eta \to \infty$ the spectrum of
Eq.(\ref{eq:eval}) is expected to be very close to the Coulomb
spectrum, $\epsilon_{n}=-1/n^{2}$, a fact, which has been used to test
the computer codes. In Fig.~\ref{fig:test_g1} it is shown that the
Coulomb limit is already well achieved for numerical values of $\eta
\geq 2.0$.  

\begin{figure}
\resizebox{0.48\textwidth}{!}{
 \includegraphics{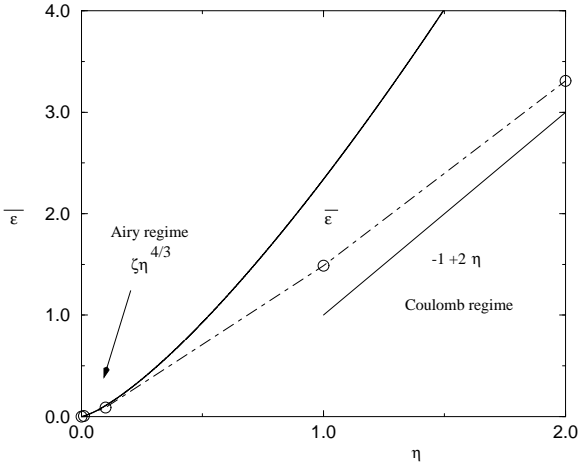}}
\caption{\label{fig:test_g2}A zoom of Fig.~\protect{\ref{fig:test_g1}}}
\end{figure} 

In the other limit, $\eta \to 0$, the spectrum of
Eqs.(\ref{eqn_solve_Bohr}) or~(\ref{eq:eval}) is expected to approach
the spectrum for a linear potential, {\it i.e.} $\epsilon_{n}=-2\eta +
\xi_{n}\eta^{4/3}$. A typical Airy-solution, however, is achieved only
for very small values,~{\it i.e.} $\eta \leq 0.02$, as
seen in Fig.~\ref{fig:test_g2}. 

For all other values, the spectrum is
somehow intermediate between those two extreme cases, as
quantitatively demonstrated in Table~\ref{eps-pred}. It is remarkable
how the curve of the calculated eigenvalues
$\bar{\epsilon}(\eta)=2\eta+\epsilon_{1}$ in Fig.~\ref{fig:test_g2}
interpolates between the two asymptotic curves $\bar{\epsilon}(1)\sim
\eta^{4/3}$ and $\bar{\epsilon}(1)\sim 2\eta -1$. 

Note that the stretching factor $z$ should be optimalized for each value of
$\eta$. The resulting value is given in Table~\ref{eps-pred} as well.
In parenthesis we note that the Coulomb limit for the present
hadronic case is reachable for $\eta \to \infty$ while for the Yukawa
case it was $\eta \to 0$ (see Section~\ref{yukawa}).    
  
\begin{table*}[t]
\begin{center}
\begin{tabular}{|l||l|c||c|c|}\hline
$\eta$&$z$ & $\bar{\epsilon}=\epsilon +2\eta$  & Airy regime & Coulomb
regime\\
& & & $\xi_1 \eta^{\frac{4}{3}}$ & $-1+2\eta $  \\\hline\hline
0.001 &0.0370 & 0.00021905 & 0.00023381 &    \\
0.01  &0.0714 & 0.00454474 & 0.00503728 &    \\ 
0.1   &0.2174 & 0.08860885 & 0.10852499 &    \\
1.0   &0.3861 & 1.48676105 & 2.3381     & 1  \\ 
2.0   &0.4762 & 3.31000318 &            & 3  \\
4.0   &0.8000 & 7.16112492 &            & 7  \\
10.0  &0.8333 & 19.04647165 &           & 19 \\\hline
\end{tabular}
\caption{\label{eps-pred} The eigenvalues $\epsilon +2\eta$ of the
  integral equation 
  (\protect{\ref{eqn_solve_Bohr}}) are given for increasing analytical
  values of 
  the physical parameter $\eta$, Column 2 gives the actual stretching
  parameter $z$. In the last two columns the corresponding eigenvalue
  for the Airy or Coulomb regime are quoted for purpose of comparison.}   
\end{center}
\end{table*}
\begin{table*}[t]
\begin{center}
\begin{tabular}{|l||c|c|c||c|c|c|}\hline
n& N=16 & N=32 & N=64 & N=16 & N=32 & N=64 \\\hline\hline
 &\multicolumn{3}{|c||}{$\eta=0.01$,$z=0.0714$}
 &\multicolumn{3}{|c|}{$\eta=0.1$, $z=0.2174$ } \\\hline 
1& 0.0045 & 0.0045 & 0.0046 & 0.0884 & 0.0886 & 0.0887 \\
2& 0.0073 & 0.0074 & 0.0075 & 0.1312 & 0.1316 & 0.1317 \\
3& 0.0092 & 0.0094 & 0.0095 & 0.1541 & 0.1548 & 0.1549 \\
4& 0.0106 & 0.0110 & 0.0110 & 0.1674 & 0.1685 & 0.1686 \\\hline\hline
&\multicolumn{3}{|c||}{$\eta=1.0$, $z=0.3861$}&\multicolumn{3}{|c|}{$\eta=10.0$, $z=0.
833$ }\\\hline\hline
1& 1.4860 & 1.4868 & 1.4871 & 19.0456& 19.0465& 19.0470\\
2& 1.8183 & 1.8193 & 1.8196 & 19.7537& 19.7551& 19.7555\\
3& 1.9090 & 1.9102 & 1.9105 & 19.8877& 19.8898& 19.8903\\
4& 1.9447 & 1.9464 & 1.9468 & 19.9339& 19.9372& 19.9379 \\\hline
\end{tabular}
\caption{\label{eival_g_c}The spectrum $\epsilon_{n}+2\eta$ for
 different values of $\eta$ and $N$, at the optimalized value of $z$.} 
\end{center}
\end{table*}
More explicit numerical results are given in App.~\ref{plots}.

\section{Summary and discussion}
\label{discussion}

Since the components of total four-momentum commute
with each other, a field theoretic Hamiltonian is formulated 
quite naturally in momentum representation.
In the instant form (usual quantization) the constituent's
degrees of freedom are the three space-like momenta 
(and their helicities and flavors),
in the front form (or in light-cone quantization) they are
the longitudinal momentum fraction $x$ and the two transversal 
momenta $\vec k _{\!\perp}$. 
Effective Hamiltonians have the same property,
they become integral equations in momentum space.
Usually, one Fourier-transforms such a momentum-space integral
equation to a Schr\"odinger-type equation in configuration space
and solves it by the familiar methods.
Taking Fourier transforms is however not always easy,
if not impossible, without additional assumptions.

In the present work we therefore want to solve the integral equations
directly in momentum space. 
In particular, we look at an integral equation with an interaction kernel like 
\[ U(\vec k,\vec k ^\prime)=-\frac{s^2}
   {(\vec k - \vec k ^\prime) ^2(c^2+ (\vec k - \vec k ^\prime)^2)}\frac{1}{\pi^2}
¸\]
as derived in Section~\ref{motivation}.
We want to calculate the eigenvalues and eigenfunctions
for the full range of the interaction parameters
$s$ and $c$, and of the physical mass $m$ of the constituent
particles, a quark and an anti-quark with equal mass.
As shown in Section~\ref{lin_pot} the solutions are a function
of only one dimensionless parameter
\[ \eta=\frac{c^3}{m s^2} ,\]
which intern can be interpreted as the ratio of two dimensionless
parameters, $c/m$ and $s/c$.
The limit $\eta\to\infty$ corresponds to a pure-Coulomb  
kernel $U\simeq (\vec k - \vec k ^\prime) ^{-2}$,
the limit $\eta\to 0$ generates the highly singular 
interaction kernel $U\simeq (\vec k - \vec k ^\prime) ^{-4}$,
as typical for a linear potential.
The latter case was also the topic of Hersbachs work \cite{hers}.

As demonstrated in several examples in Section~\ref{combi},
the transitional region with $\eta \sim 1.0$ 
corresponds to a superposition of a Coulomb and a Yukawa potential,
which both are studied on their own merit in 
Sections~\ref{coulomb} and \ref{yukawa}, respectively.
Therefore, for values of $s$ and $c$
fixed by their default value in Eq.(\ref{s.c.def}),
the spectra for a mass larger than typically 180 GeV are 
more like those in a linear potential, as opposed
to more Coulomb-like spectra for masses smaller than
typically 1.8 GeV. Masses in between have led to a mixed pattern.

The technical problem of solving the interaction kernel
in momentum space is approached by discretization
via Gaussian quadratures, and diagonalization
of the so generated Hamiltonian matrix.
Much attention is paid to speed up convergence
by a counter-term technology already developed 
for the pure Coulomb case \cite{kpw92,wdipl}.
It is summarized in Sections~\ref{gaussian} and \ref{coulomb}.
In Sections~\ref{yukawa} and~\ref{combi} it is adapted
to the Yukawa and the combined Yukawa plus Coulomb problem,
in this work refered to as the hadronic Coulomb problem by obvious reasons.
Special emphasis is put on a free formal parameter,
called the stretching factor $z$, which can be adjusted
for a considerably increased numerical precision
and stability. This way, one can restrict -- on the average -- 
on matrix diagonalization problems with a matrix dimension
as small as $N=32$.
This part of the technical problem is applicable
to many other physical problems.
The restriction to equal masses of the constituents
can be relaxed easily.

We find it remarkable, that we solve a problem
which on the technical level looks like a problem
of usual (equal-usual-time) quantization
despite the fact that the generated solutions
hold for the front form.
The equation actually being solved cannot be
recognized as to have its roots in light-cone quantization.
The only physical assumption entering the considerations
in Section~\ref{motivation}
is that the current-current term was replaced by
the term of leading order in Eq.(\ref{current}).

It is, of course, still a long way to go for solving
an effective QCD Hamiltonian on the technical level.
Only a few steps have been taken in the present work.
On the long run, we want to 
include properly the non-local factors $A(\vec k)$ 
in Eq.(\ref{start}), to relax the assumption
of Eq.(\ref{current}), and to insert more general
expressions for the effective coupling constant
$\bar{\alpha}(Q)$.
Work in this direction is under way.



\begin{appendix}
\section{Some selected numerical results}
\label{plots}
In this appendix we add a few typical results for the hadronic Coulomb
potential as discussed in Section~\ref{combi}. The spectra and s-wave
eigenfunctions for a very small and a  
large value of $\eta$ are shown to illustrate the main
differences. The structure of the spectrum for $\eta=0.01$ is similar
to those of a linear potential. With increasing values of $\eta$, the
shape of the spectrum changes to that of a Coulomb potential
particularly for the large value $\eta=10.0$. 
More results are available from the authors on request.

\begin{figure}[ht!]
\resizebox{0.48\textwidth}{!}{ 
\includegraphics{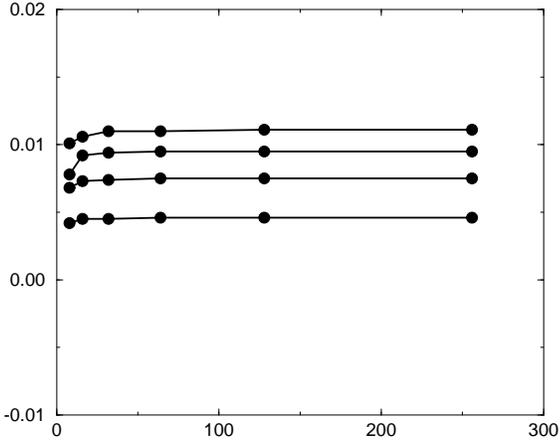}}
\caption{\label{fig:12}
   The eigenvalues $\epsilon +2\eta$ for $\eta=0.01$ ($z=0.0714$) are
   plotted versus the   
   number of integration points $N$~($8,16,32,64,128,256$).- Note the
   almost equidistant structure as in a linear potential.}
\end{figure}
\begin{figure}[ht!]
\resizebox{0.48 \textwidth}{!}{ 
\includegraphics{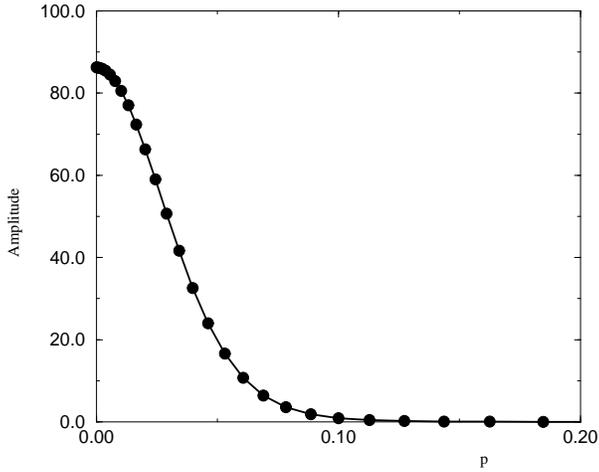}}
\caption{
   \label{fig:16}
   The s-wave eigenfunction for $\eta=0.01$ is plotted versus the
   momenta in Bohr units.- Note that the
   most of the integration points are in that region where the wave
   function is significantly different from 
   zero. This is due to the choice of $z=0.0714$.}
\end{figure}

\begin{figure}[ht!]
\resizebox{0.48\textwidth}{!}{ 
\includegraphics{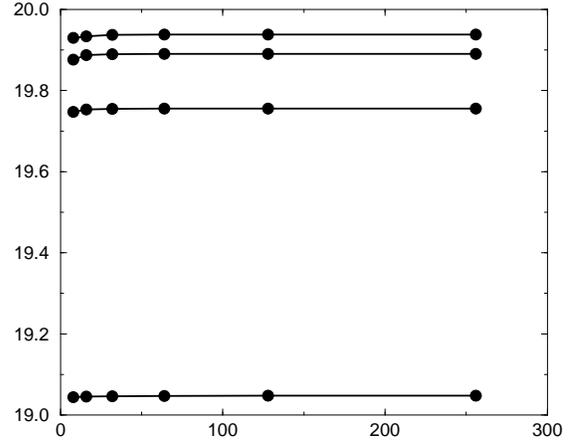}}
\caption{\label{fig:15} 
   The eigenvalues $\epsilon +2\eta$ for $\eta=10.0$ ($z=0.833$) are
   plotted versus the same  
   number of integration points as left.- 
   This spectrum is of the Coulomb type, see Fig.~\protect{\ref{fig:2}}.}   
\end{figure} 

\begin{figure}[ht!]
\resizebox{0.48\textwidth}{!}{
\includegraphics{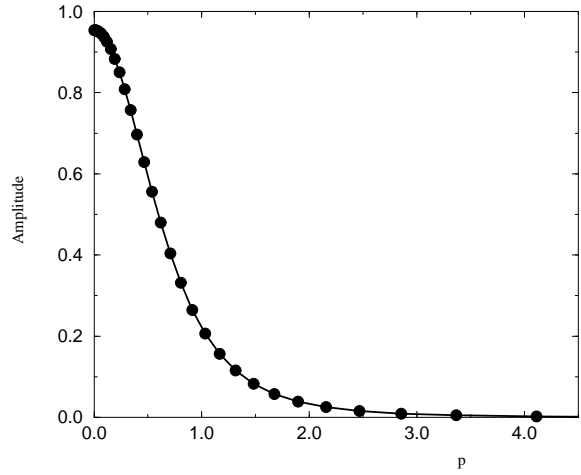}}
\caption{\label{fig:19}
The s-wave eigenfunction for $\eta=10.0$ ($z=0.833$) is plotted versus
the momentum values $p/p_{Bohr}$.- This figure is comparable with
Fig.\protect{\ref{fig:3}}.}
\end{figure}
The following parts are not included in the printed work.
\section{Results for the Yukawa potential}
\begin{figure}[h!]
\resizebox{0.48\textwidth}{!}{
\includegraphics{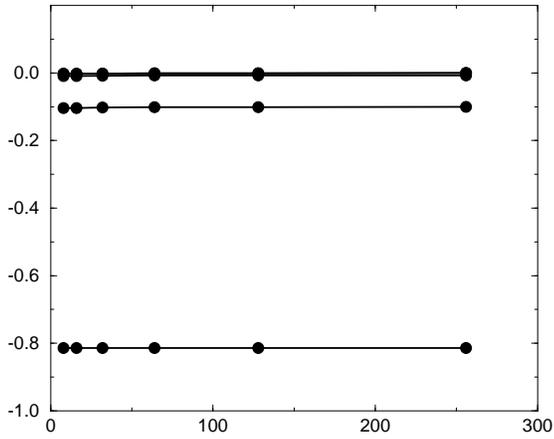}}
\caption{\label{fig:Y1}Spectrum for $\eta=0.01$}
\end{figure}
\begin{figure}[h!]
\resizebox{0.48\textwidth}{!}{
\includegraphics{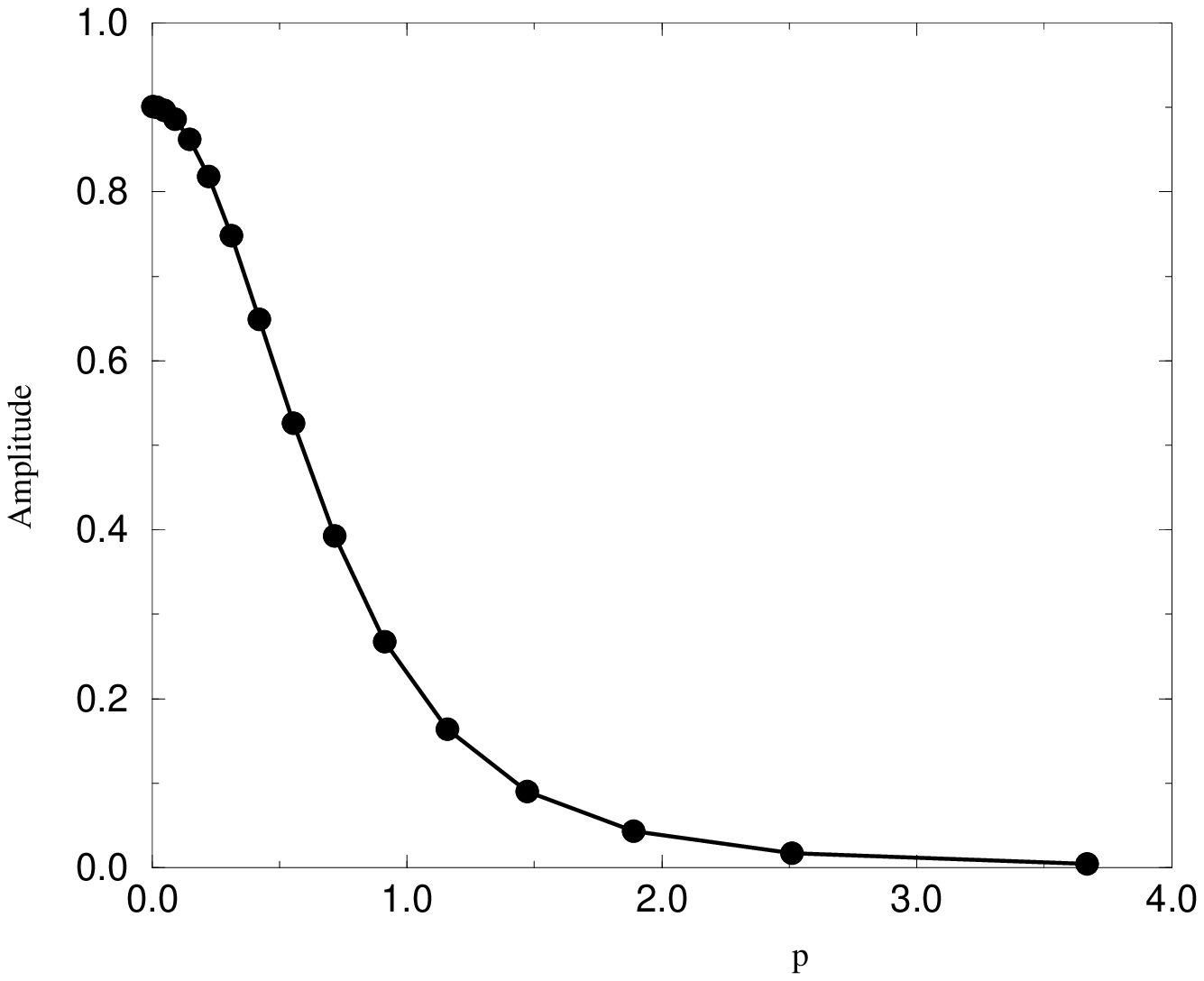}}
\caption{\label{fig:Y2}S-wave eigenfunction for $\eta=0.01$}
\end{figure}
\begin{figure}[ht!]
\resizebox{0.48\textwidth}{!}{
\includegraphics{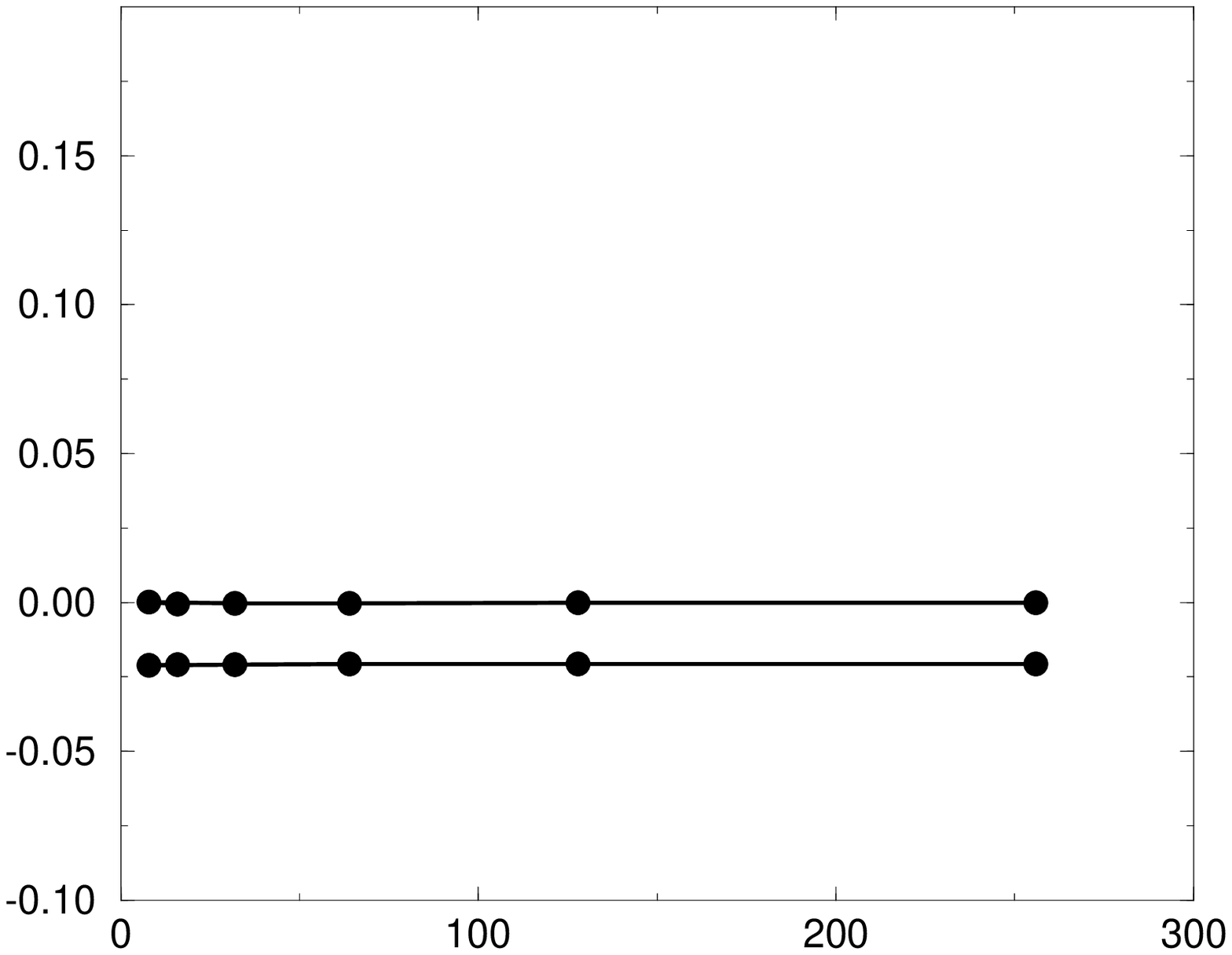}}
\caption{\label{fig:Y3}Spectrum for $\eta=1.0$}
\end{figure}
\begin{figure}[ht!]
\resizebox{0.48\textwidth}{!}{
\includegraphics{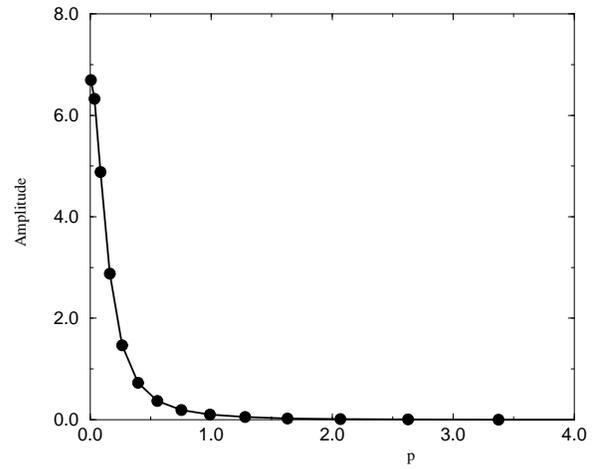}}
\caption{\label{fig:Y4}S-wave eigenfunction for $\eta=0.1$}
\end{figure}
\clearpage
\newpage
\section{more numerical results for the hadronic Coulomb potential}
\begin{figure}[h!]
\resizebox{0.48\textwidth}{!}{
\includegraphics{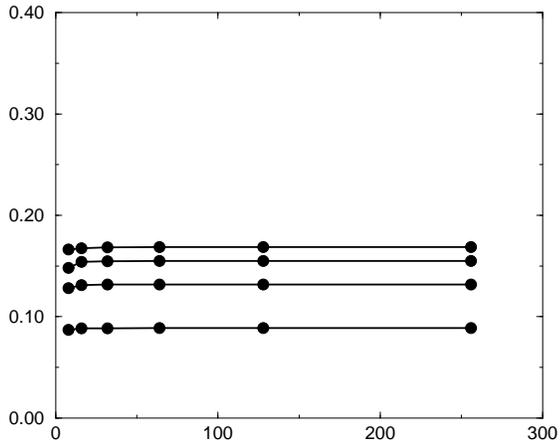}}
\caption{\label{fig:C1}Spectrum for $\eta=0.01$}
\end{figure}
\begin{figure}[h!]
\resizebox{0.48\textwidth}{!}{
\includegraphics{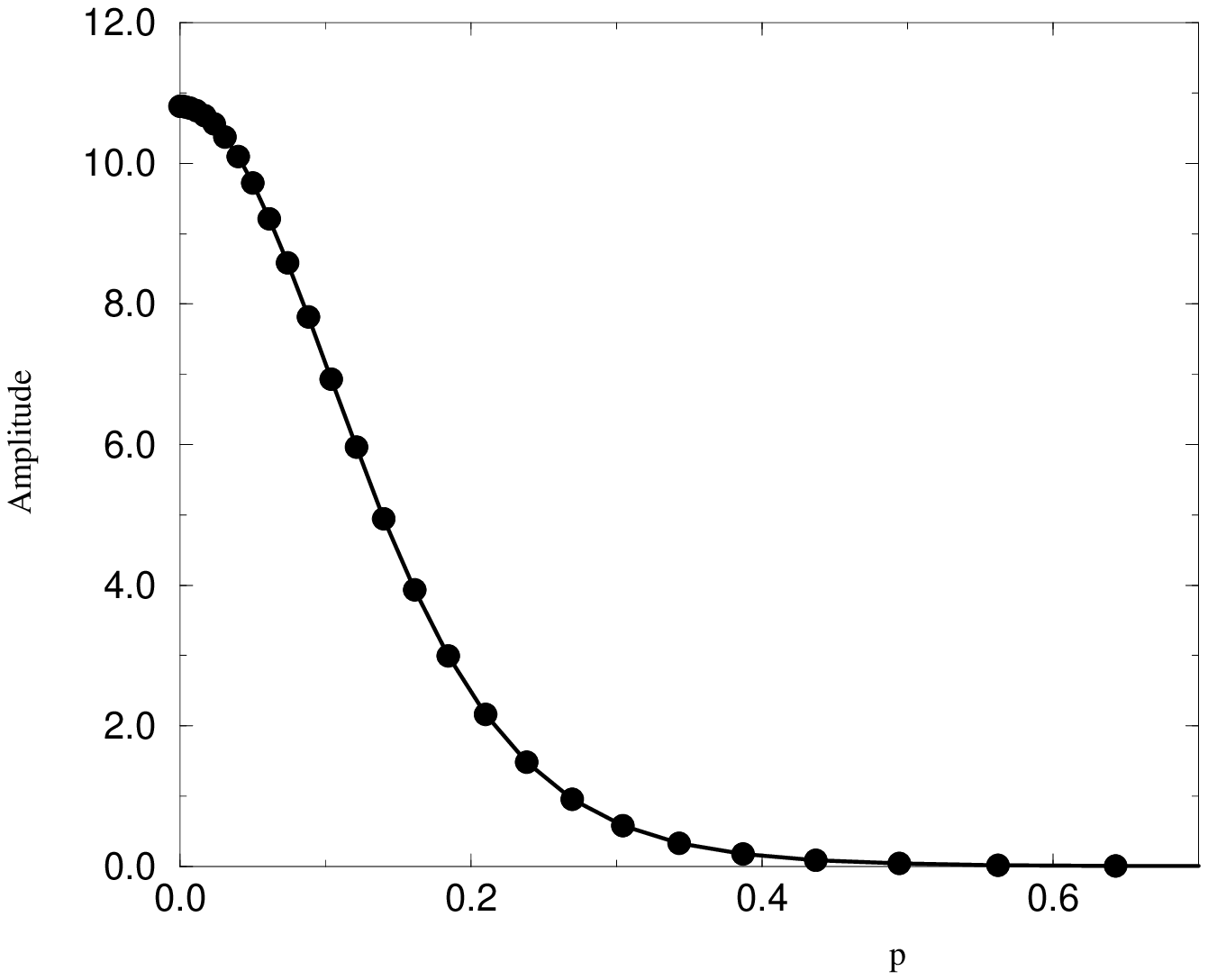}}
\caption{\label{fig:C2}S-wave eigenfunction for $\eta=0.1$}
\end{figure}
\begin{figure}[ht!]
\resizebox{0.48\textwidth}{!}{
\includegraphics{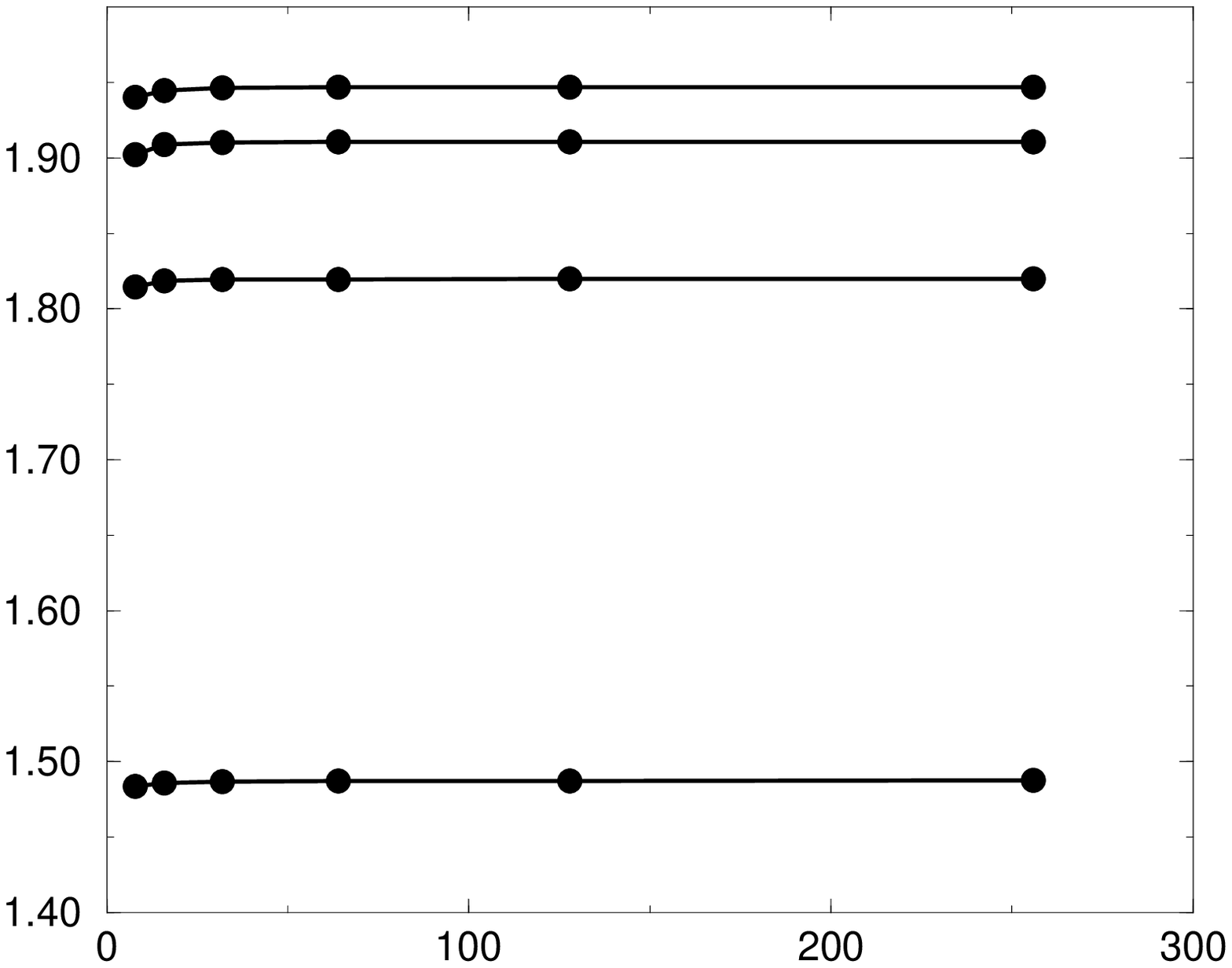}}
\caption{\label{fig:C3}Spectrum for $\eta=1.0$}
\end{figure}
\begin{figure}[ht!]
\resizebox{0.48\textwidth}{!}{
\includegraphics{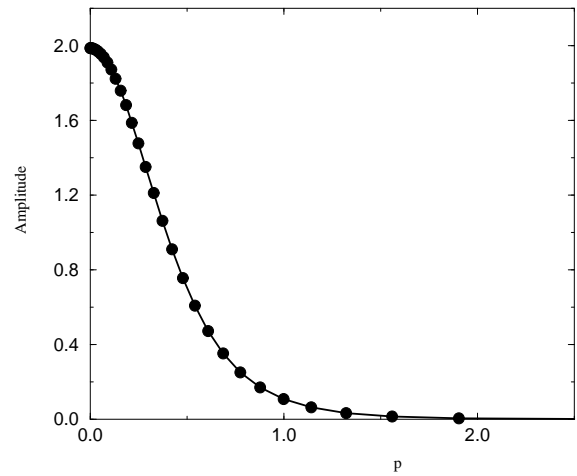}}
\caption{\label{fig:C4}S-wave eigenfunction for $\eta=1.0$}
\end{figure}
\clearpage
\newpage
\section{Determination of $z$ for the hadronic Coulomb potential}
\begin{figure}[h!]
\resizebox{0.48\textwidth}{!}{
\includegraphics{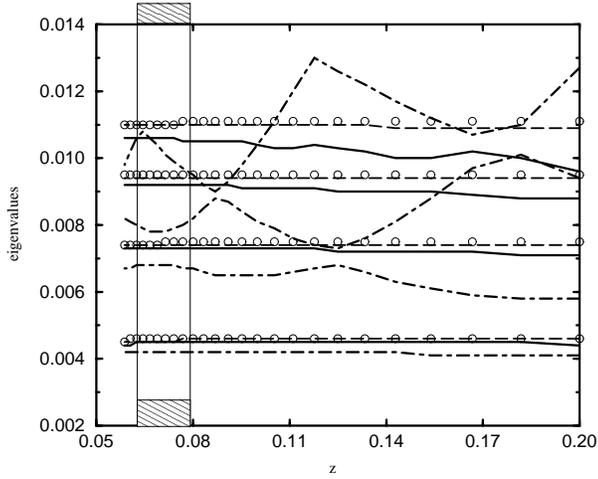}}
\caption{\label{fig:Z1}Eigenvalues for $\eta=0.01$ versus stretching $z$}
\end{figure}
\begin{figure}[h!]
\resizebox{0.48\textwidth}{!}{
\includegraphics{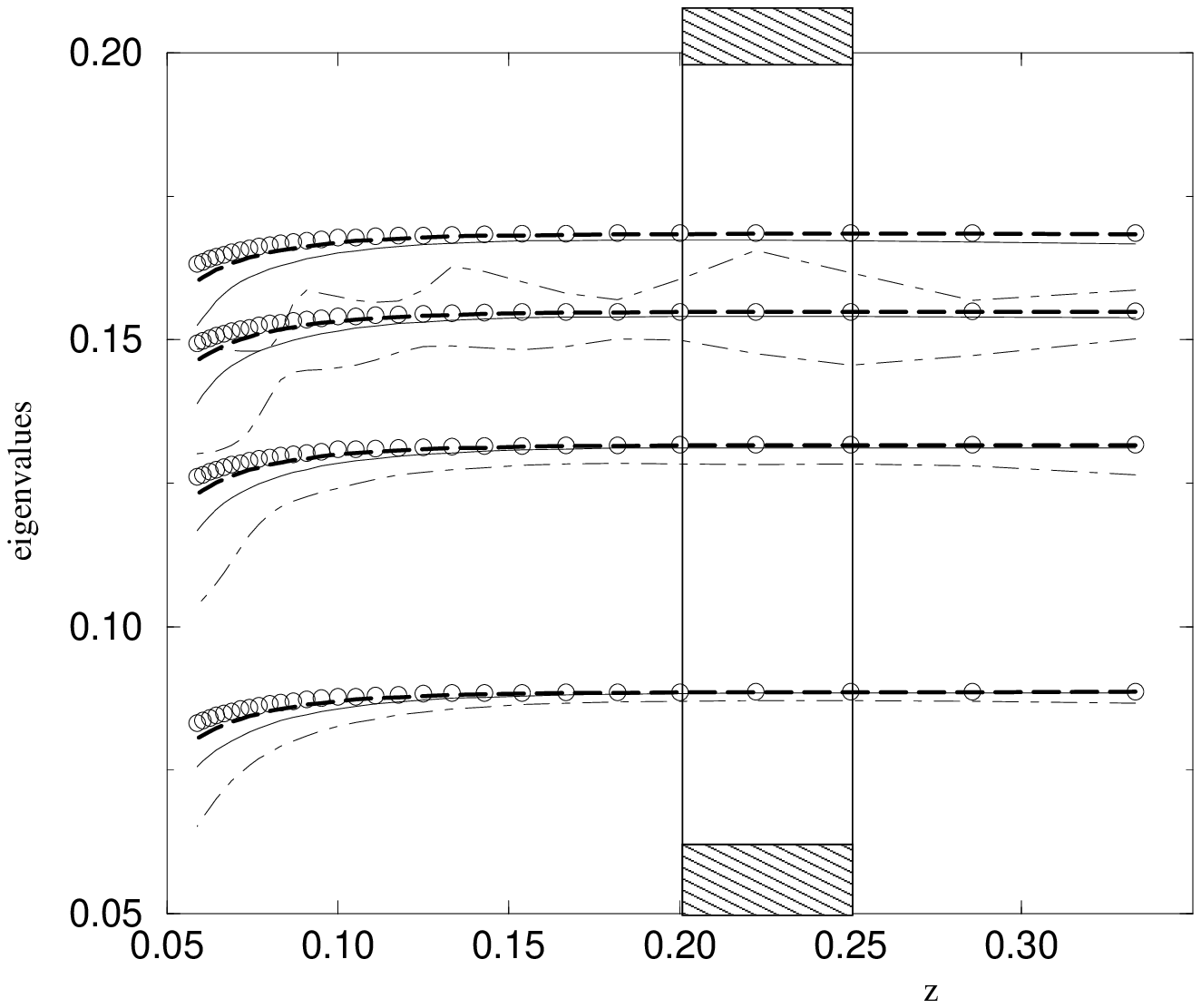}}
\caption{\label{fig:Z2}Eigenvalues for $\eta=0.1$ versus stretching $z$}
\end{figure}
\begin{figure}[ht!]
\resizebox{0.48\textwidth}{!}{
\includegraphics{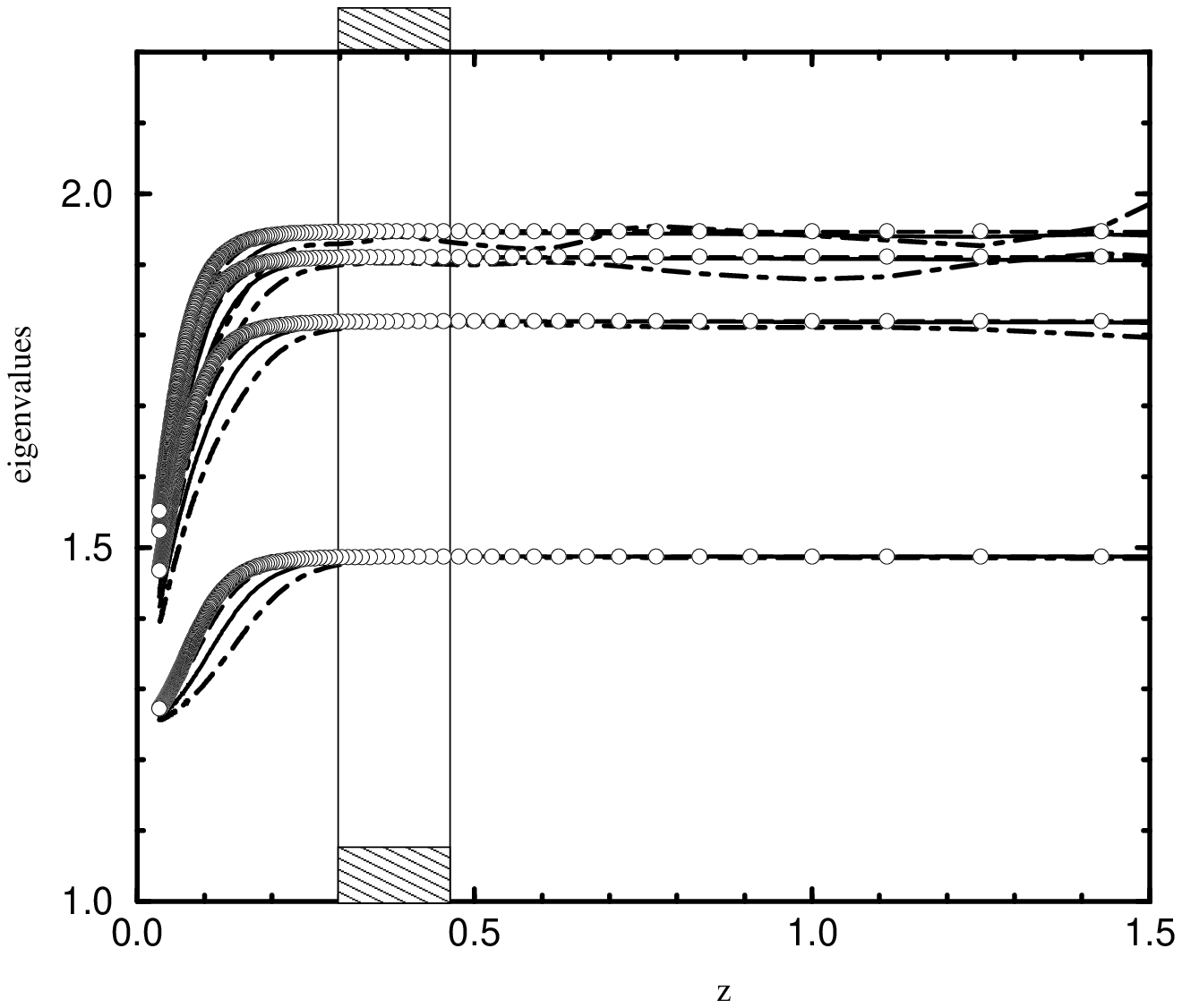}}
\caption{\label{fig:Z3}Eigenvalues for $\eta=1.0$ versus stretching $z$}
\end{figure}
\begin{figure}[ht!]
\resizebox{0.48\textwidth}{!}{
\includegraphics{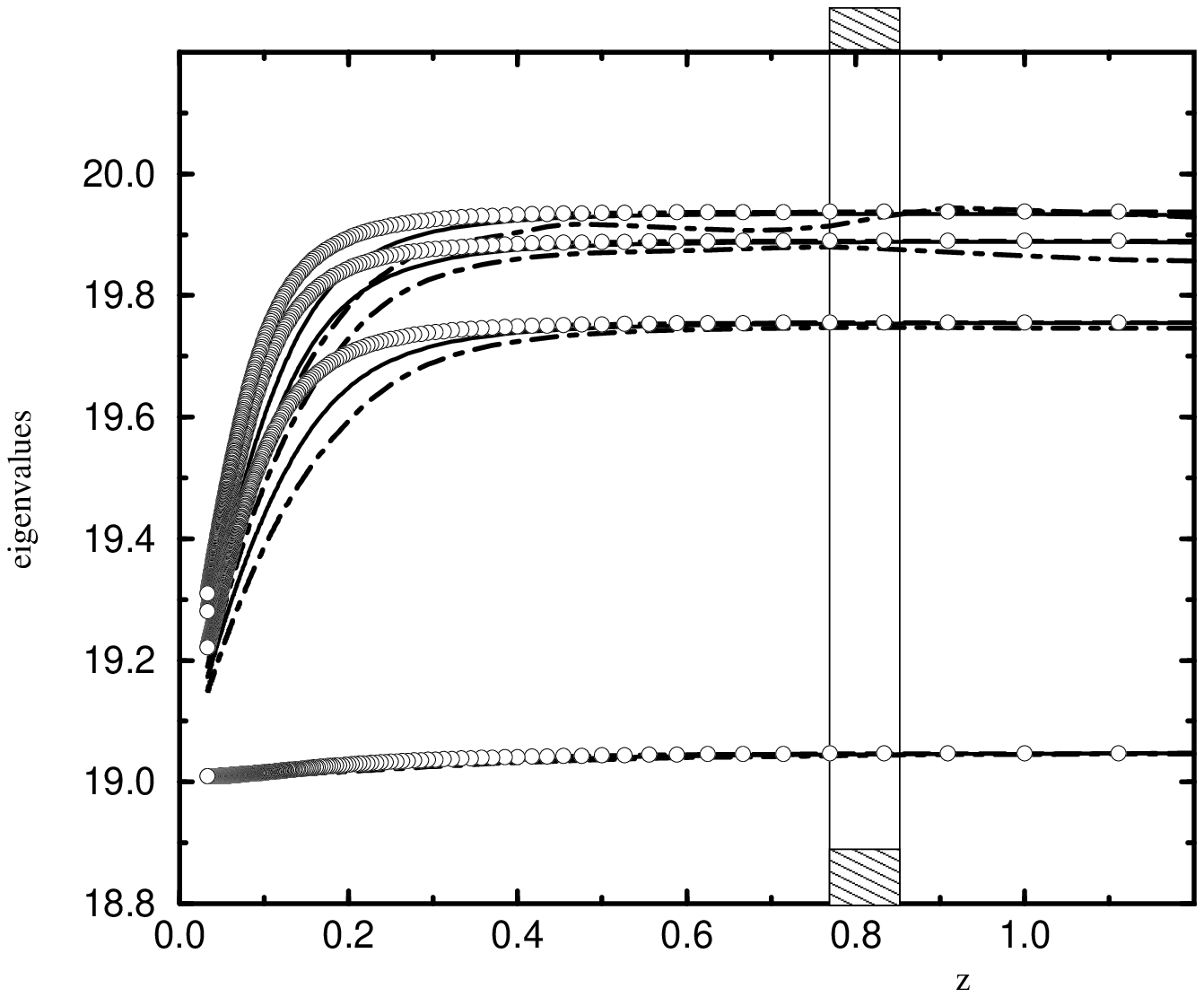}}
\caption{\label{fig:Z4}Eigenvalues for $\eta=10.0$ versus stretching $z$}
\end{figure}


\end{appendix}   
\end{document}